
\input phyzzx
\hoffset=0.2truein
\voffset=0.1truein
\hsize=6truein
\def\TITLEPAGE{\frontpagetrue}
\def\CALT#1{\hbox to \hsize{\tenpoint \baselineskip=12pt
        \hfil \vtop{
        \hbox{\strut CALT-68-#1}
        \hbox{\strut DOE RESEARCH AND}
        \hbox{\strut DEVELOPMENT REPORT}}}}
\def\CALTECH{
        \address{California Institute of Technology,
    Pasadena, CA 91125\foot{{\rm address after 1 September: Institute for
Advanced Study, Princeton, NJ 08540}}}}
\def\TITLE#1{\vskip.5in \centerline{\fourteenpoint#1}}
\def\AUTHOR#1{\vskip.2in \centerline{#1}}
\def\ABSTRACT#1{\vskip.2in \vfil \centerline
            {\twelvepoint \bf Abstract}
                     #1 \vfil}
\def\ENDTITLEPAGE{\vfil \eject \pageno=1}
\hfuzz=5pt
\tolerance=10000
\TITLEPAGE
\CALT{1921}
\TITLE{Scattering from Electroweak Strings
\foot{This work supported in part
by the U.S. Department of Energy under Grant No.
DE-FG03-92-ER40701}}
\AUTHOR{Hoi-Kwong Lo\foot{HKL@THEORY.CALTECH.EDU}}
\CALTECH
\ABSTRACT{The scattering of a charged fermion from an electroweak string
is studied. Owing to an amplification of the wave function at the core
radius, the cross sections for helicity flip processes can be largely
enhanced. For $0 <\sin^2 \theta_w < 1/2 $ (where $\theta_w$ is the Weinberg
angle), $\omega \sim k \sim m_e$ and $kR \ll 1$, we show
that the helicity flip differential
cross section for electrons is of the order $m_e^{-1}$ and is independent
of angle. We compare our results with those obtained in
calculations of rates for baryon number violating processes
in the core of a cosmic string. In that case,
while the enhancement is a general phenomenon, its actual magnitude is
extremely sensitive to the fractional flux carried by
the string core. Apart from showing the existence of a similar enhancement
effect for non-topological strings,
our results indicate that in some models
the magnitude of enhancement
can be rendered much less sensitive to
the value of the parameters in the theories. Scattering of particles off
semi-local strings and axion strings are also considered.}
\ENDTITLEPAGE
\eject

\def\gmu{\gamma^{\mu}}
\def\al{\alpha_L}
\def\ar{\alpha_R}
\chapter{Introduction}
Some years ago, Callan\Ref\Cal{C. Callan, Phys. Rev.
{\bf D25}, 2141 (1982); Phys. Rev. {\bf D26}, 2058 (1982); Nucl. Phys. {\bf
B212}, 391 (1983).}
and Rubakov\Ref\Rub{V. Rubakov, JETP Lett. {\bf 33},
644 (1981); Nucl. Phys. {\bf B203}, 311 (1982).}
(see also Wilczek\Ref\Wil{F. Wilczek, Phys. Rev. Lett. {\bf 48}, 1146 (1982).})
showed that a grand unified monopole may catalyze baryon number violating
processes with strong interaction cross sections rather than the
much smaller geometric cross sections. This enhancement effect can
be understood as a consequence of a large amplification of the
fermionic wave functions near the location of the monopole.\Ref\Bran{R. H.
Brandenberger, A.-C. Davis, and A. Matheson, Nucl. Phys. {\bf B307},
909 (1988); R. H. Brandenburger  and L. Perivolaropoulos, Phys. Lett.
{\bf B208}, 396 (1988).}

A similar enhancement of cross section also occurs for cosmic strings with
fractional fluxes:\REF\Alf{M. G. Alford, J. March-Russell, and F. Wilczek,
Nucl. Phys. {\bf B328}, 140 (1989).}\REF\Per{W. B. Perkins, L.
Perivolaropoulos, A.-C. Davis, R. H. Brandenberger, and A. Matheson,
Nucl. Phys. {\bf B353}, 237 (1991).}\refmark{\Alf,\Per}
Alford, March-Russell, and Wilczek\refmark{\Alf}
studied the fermion number violating
process in a cosmic string core due to a Yukawa coupling. In their model,
there are two fermions  with equal $U(1)$ charges and two scalar fields.
The first scalar field, $\eta$, acts as the Higgs field and thus condenses
outside string core. The second scalar field, $\phi$, which has a Yukawa
coupling to the two fermion fields, condenses within the core.
In the limit $kR  \ll \nu R \ll 1$, where $k$ is the momentum of the
incident fermion,
$R$ is the size of the core and $\nu= \lambda < \phi >$ ($\lambda$ being
the Yukawa coupling constant), they found generic enhancement by large
factors over the naive, geometric cross section. Maximal enhancement occurs
when ${d \sigma \over d \theta} \sim { 1 \over k}$.

A prominent feature of their results is that while a large enhancement
of the fermion number violating process is a general phenomenon, its actual
magnitude is extremely sensitive to the $U(1)$ charge of the fermions. For
instance, changing the charge from $\alpha = 1/2 $ to $\alpha = 1/4$ results
in a diminuation of the inelastic cross section by around 15 orders of
magnitude. By assigning baryon numbers to the fermions and scalars,
their results imply that the exact magnitude of the baryon number violating
process is very sensitive to the details of the grand unification model
under consideration.
Since there are uncertainties in our experimental
determination of low energy parameters such as the Weinberg angle,
it might be hard for us to say for sure whether a model is phenomenologically
feasible. The point is that
a slight error made in our determination of the values of such parameters
leads to a huge variation in the rate of baryogenesis and
may render a feasible model unfeasible and vice versa. For this reason,
we would like to ask the following question: Is this sensitivity a
generic feature or is it model-dependent? In other words, can we
construct a model where the inelastic cross section is less sensitive
to the values of the parameters?

A hint to the answer to this question comes from the investigation made
by Perkins {\it et al.}\refmark{\Per}
In their paper, the cross section
for a baryon number violating process was derived using first order
perturbation theory in quantum field theories. The transition matrix element
between an initial state $\ket{\psi}$ and a final state $\ket{\psi'}$ is given
by
${\cal A} = \langle \psi'| \int d^4 x {\cal L}_I (x) \ket{\psi}.$
The computation was divided into two steps. Firstly, they evaluated ${\cal A}$
using free fermion spinors, resulting in the ``geometric'' cross section.
In the second step, they solved the Dirac equation with
the appropriate boundary
conditions to determine the amplitude of the spinor at the core radius R and
defined the amplification factor $A$ as the ratio of the amplitude of the
actual spinor to that of a free spinor. Since the cross section is
proportional to ${\cal A}^2$ and since ${\cal A}$ involves two spinors, the
catalysis cross section is enhanced by a factor $A^4$ over the geometric
cross section. This argument relies on the fact that the amplification factor
for the initial state is the same as that for the final state.
It might
be possible for us to construct models with different amplification factors
for the initial and final states. If the two amplifications have opposite
dependence on the parameters, the overall amplification, which is
the product of the two amplification factors, will then be insensitive
to the parameters in the model.

In this paper, we study the scattering of a charged fermion from an
electroweak string. We show that for $0 < \sin^2 \theta_w < 1/2$
($\theta_w=$Weinberg angle), $\omega \sim k \sim  m_e $ and
$kR \ll 1$, the helicity flip differential cross section for electrons is
of the order $m_e^{-1}$. A delicate cancellation of the dependence of
the two amplification factors on the Weinberg angle indeed occurs within
this regim\'e.
We would like to remark that the differential cross section in this regim\'e
is dominated by a single partial wave and is thus
independent of angle.

Incidentially, our results illustrate that
the analysis of the enhancement effects for
cosmic strings can be extended
to a wider class of string defects: the semi-local
strings\Ref\Vac{T. Vachaspati and A. Ach\'ucarro, Phys. Rev. {\bf D 44},
3067 (1991).}
and the electroweak strings.\Ref\Va{T. Vachaspati, Phys. Rev. Lett. {\bf
68}, 1977 (1992).}
These recently discovered defects
occur in theories where the fundamental group of the vacuum manifold
is trivial. Thus, they are at best metastable.\Ref\Hin{M. Hindmarsh,
Phys. Rev. Lett. {\bf 68}, 1263 (1992); M. James, L. Perivolaropoulos,
and T. Vachaspati, Phys. Rev. {\bf D 46}, 5232 (1992); Nucl. Phys. {\bf
B 395}, 534 (1993).}
While Z-flux carrying electroweak strings are unstable in the Weinberg-Salam
model, various mechanisms for stabilizing a Z-flux string have been
proposed. First, one can add a linear time dependence of the Goldstone boson
to obtain a stable spinning vortex solution.\Ref\Perivo{L. Perivolaropoulos,
hep-ph/9403298.}
Second, fermions that are massive outside the core become massless inside.
It is clear that there are superconducting zero modes in the core.
These bound states tend to stabilize the non-topological
solitons.\Ref\Vachas{T. Vachaspati and R. Watkins, Phys. Lett. {\bf B318},
163 (1993); T. D. Lee, Phys. Rep. {\bf 23C}, 254 (1976).}
Another
possibility would be to consider extensions of the electroweak model
or topological strings carrying Z-flux which are formed in an earlier
phase transition.\Ref\Dva{G. Dvali and G. Senjanovi\'c, hep-ph/9403277;
R. Brandenburger, A.-C. Davis, and M. Trodden, hep-ph/9403215.}

Baryogenesis during the weak
phase transition is particularly interesting
as it may eventually be experimentally verifiable.
Consider the following wild speculative scenario: non-topological
electroweak strings are formed at the electroweak phase transition. They
are stabilized by some mechanism (either one of the above or a combination
or some other means). Baryogenesis occurs inside their cores.
Baryogenesis due to electroweak strings in the two-Higgs
model
has been discussed in the literature.\Ref\Bar{See M. Barriola,
hep-ph/9403323 and references therein.}
It would be interesting to understand the relevance of
our results to baryogenesis in future investigations.

After the completion of an earlier version of this paper, we
received a revised manuscript by Davis, Martin and Ganoulis\Ref\Dav{A.-C.
Davis, A. P. Martin, and N. Ganoulis, Cambridge Preprint DAMTP 93-46.}
which also discussed electrons scattering off electroweak strings
for $0 \leq \sin^2\theta_w \leq 1/2 $ and $ k \ll m$ or
$k \gg m$. In this paper, we consider the whole parameter space
$0 \leq \sin^2\theta_w \leq 1 $ in the regim\'e $m \sim k$.
In particular, our analysis applies to semi-local strings, which
correspond to $ \sin^2\theta_w = 1 .$

The plan of this paper is as follows. In section 2, we review
the subject of the electroweak strings and describe a simple model
of the field configuration that we will be working with.
Using a partial wave analysis,
the differential cross section for the helicity flip
process of electrons for various values of $\theta_w$ will be computed
in section 3. In particular, we show that the cross section is proportional
to $m_e^{-1}$ for $ 0<\sin^2 \theta_w <1/2$, $\omega \sim k \sim m_e$
and $k R \ll 1 $.
Moreover, the result for the semi-local strings can be obtained from that of
the electroweak strings by setting $\sin^2 \theta_w $ to $1$.
In our concluding remarks in section 4, we also note that helicity
is violated outside the core of an axion string. Thus, it makes no sense
to discuss helicity conserving and helicity flip cross sections
in this context.

\chapter{Extended Abelian Higgs Model and Electroweak Strings}

Consider an extension of the Abelian Higgs model with $N=2$ complex
scalars $\Phi$ with their overall phase gauged and an $SU(2)$ global
symmetry. The most general renormalizable Lagrangian in four dimensions
consistent with these symmetries is
$$
{\cal L} = | D_\mu \Phi |^2 - {1 \over 2} \lambda \bigl( | \Phi|^2
- \eta^2 \bigr )^2 - {1 \over 4} F_{\mu \nu} F^{\mu \nu}. \eqno(1)$$
The field $\Phi$ acquires a vacuum expectation value of magnitude
$\eta$ and the symmetry is spontaneously broken into a global $U(1)$.
It has also been shown that the Nielsen-Olesen vortex solutions\Ref\Nie{H.
B. Nielsen and P. Olesen, Nucl. Phys. {\bf B 61}, 45 (1973).}
of the Abelian Higgs model (the case with N=1) carry over to the extended
Abelian Higgs model. However, the stability of such vortex solutions
becomes a dynamical question and depends on the ratio of the masses of
the Higgs and vector particles.\refmark{\Hin}

Now the extended Abelian Higgs model is precisely the Weinberg-Salam
model\Ref\Wei{S. Weinberg, Phys. Rev. Lett. {\bf 19}, 1264 (1967); A.
Salam, in {\it Proceedings of the Eighth Nobel Symposium,} ed. N.
Svartholm (Almqvist and Wiksell, Stockholm, 1968).}
with the $SU(2) $ charge set equal to zero. By gauging the $SU(2)$
symmetry, one obtains string solutions in the electroweak theory. Such
electroweak strings are non-topological and unstable in the minimal
electroweak theory. They  may, however, be made metastable in some
extended models.

Consider an electron moving in the background field of an electroweak
string. The relevant part of the Lagrangian is
$$
{\cal L} = i \bar L \gmu D_\mu L +i \bar e_R \gmu D_\mu e_R-
f_e( \bar L e_R \Phi + \Phi^{\dag} \bar e_R L), \eqno(2)$$
where $\bar L=(\bar \nu, \bar e_L) $, $f_e$ is the Yukawa coupling
constant, $\Phi$ is the usual Higgs doublet, and the covariant
derivative has the form
$$ D_\mu = \partial_\mu +{i \alpha \gamma \over 2} Z_\mu , \eqno(3)$$
where $\gamma=e/(\sin \theta_w \cos \theta_w) $ ($\theta_w $ being the
Weinberg angle) and the $Z-$coupling, $\alpha$, is given by\Ref\Pre{J.
Preskill, Phys. Rev. {\bf D46}, 4218 (1992).}
$$ \alpha= -2 (T_3 - Q \sin^2 \theta_w ), \eqno(4)$$
where $T_3$ is weak isospin and $Q$ is electric charge. Note that for
electrons and down quarks,
$$\alpha_L= \alpha_R +1, \eqno(5)$$
and there is a marked asymmetry between left and right fields.

For explicit calculations, consider the following simple model\refmark{\Dav}
of the field configuration.
$$\eqalignno{ \Phi &=\left( \matrix{\Phi^+ & \Phi^0 \cr} \right)=
 f(r) e^{i\theta}\left({\matrix{ 0\cr 1\cr
}}\right)\cr  Z_\phi &= -{v(r)/r}\cr
             Z_r &=W=A=0 \cr
             f(r) &=\cases{0   &$r<R$\cr
                           {\eta \over \sqrt 2} &$r>R$\cr}\cr
             v(r) &=\cases{0   &$r<R$\cr
                           {2 \over \gamma} &$r>R$\cr} &(6) \cr }$$
where $Z$ and $W$ are the gauge bosons and $A$ is the photon field.
We expect our results to be insensitive to the detail of the core model.
A discussion about this issue can be found in Ref. 6.

Writing $e_L= \left( \matrix{0 \cr \psi \cr } \right)$ and $e_R=
\left( \matrix{\chi \cr 0 \cr} \right)$, in the representation
$$\gamma^0= \left( \matrix{0 & 1 \cr 1 & 0 \cr} \right),
 \gamma^i= \left( \matrix{0 & -\sigma^j \cr \sigma^j & 0 \cr}\right) ,
 \gamma^5= \left( \matrix{1 & 0 \cr 0 & -1 \cr} \right),\eqno(7)$$ the
Hamiltonian is
$$
H=\left( \matrix{ -i \sigma^j D^R_j & f_e f e^{-i \theta} \cr
            f_e f e^{i \theta} & i \sigma^j D^L_j  \cr} \right).
\eqno(8)$$
The equations of motion for $\psi$ and $\chi$ are
$$ \eqalign{ \omega \chi &+ i \sigma^j D^R_j \chi\cr
               \omega \psi &- i \sigma^j D^L_j \psi\cr}
\eqalign{- \cr - \cr}
\eqalign{ f_e f e^{-i \theta}
               \psi &=0 \cr
  f_e f e^{i \theta}
               \chi &=0.\cr} \eqno(9)$$
Note the phase $e^{i \theta}$ and the coupling of $\psi$ to $\chi$ via
the mass term. Inside the core, there is no coupling and electron is
massless. The helicity operator is given by
$$ \Sigma \cdot \Pi = \left( \matrix{ \sigma \cdot \pi_R & 0 \cr
                                     0 & \sigma \cdot \pi_L \cr} \right)
      =\left( \matrix{ -i \sigma^j D^R_j & 0  \cr
             0  & -i \sigma^j D^L_j \cr } \right). \eqno(10)$$
To see that helicity is not conserved, we compute its commutator with
the Hamiltonian and find it to be non-zero inside the core.\Ref\Gan{N.
Ganoulis, Phys. Lett. {\bf B298}, 63 (1993).}
$$ [ H, \Sigma \cdot \Pi ]= if_e \left( \matrix{0 & \sigma^j
(D_j \Phi^0)^* \cr \sigma^j D_j \Phi^0 & 0\cr } \right). \eqno(11) $$
Note that helicity violating processes can only occur in the string core.
They can, however, be enhanced by an amplification of the fermionic wave
function at the core radius. In the following section, we perform a detailed
calculation of the differential cross section for such scattering processes.

\chapter{Scattering Amplitude}
We try the usual partial wave decomposition.
$$
\eqalignno{\chi (r, \theta) &= \sum_{l=- \infty}^\infty \left(
            \matrix{ \chi^l_1(r) \cr i \chi^l_2(r) e^{i \theta} \cr} \right)
           e^{il \theta} \cr
           \psi (r, \theta) &= \sum_{l= - \infty}^\infty \left(
            \matrix{ \psi^l_1(r) \cr i \psi^l_2(r) e^{i \theta} \cr} \right)
           e^{i(l+1) \theta}. & (12)  \cr } $$
Making use of
$$ \sigma^j D_j= \left( \matrix{0 & e^{-i \theta} (D_r-i D_\theta) \cr
                               e^{i \theta} (D_r +i D_\theta) & 0 \cr }
                   \right), \eqno(13) $$
we substitute (12) into (9) to obtain

$$
\matrix{\omega \chi^l_2 &+&\hfill \left( {d \over dr} -{l \over r} +
      {\ar \gamma \nu \over 2r} \right) \chi^l_1 & - f_e f \psi^l_2 &= 0 \cr
           \omega \chi^l_1 &-& \left( {d \over dr} +{ l+1 \over r}-
 {\ar \gamma \nu \over 2r} \right) \chi^l_2 & - f_e f \psi^l_1 &=0 \cr
           \omega \psi^l_2 &-&\left( {d \over dr} -{ l+1 \over r }+
 {\al \gamma \nu \over 2r} \right) \psi^l_1 & - f_e f \chi^l_2  &=0 \cr
           \omega \psi^l_1 &+&\left( {d \over dr} +{ l+2 \over r }-
 {\al \gamma \nu \over 2r} \right) \psi^l_2 & - f_e f \chi^l_1  &=0.\cr}
  \eqno(14)$$

\noindent{\bf (a) Internal Solution ($r < R$)}

In this region, $f=v=0$, so the equations of motion (14) reduce to
$$
\matrix{\omega \chi^l_2& +&\hfill \left( {d \over dr} -{l \over r}
  \right) \chi^l_1 &= 0 \cr
           \omega \chi^l_1& -& \left( {d \over dr} +{ l+1 \over r}
  \right) \chi^l_2 &=0 \cr
           \omega \psi^l_2& -&\left( {d \over dr} -{ l+1 \over r }
  \right) \psi^l_1 &=0 \cr
           \omega \psi^l_1& +&\left( {d \over dr} +{ l+2 \over r }
  \right) \psi^l_2 &=0 .\cr} \eqno(15)$$
Thus, $\psi$ and $\chi$ are decoupled from each other in the string core.
Combining the first two equations and setting $z=\omega r$, we obtain
$$
{1 \over z} {d \over dz} \left( z{d \over dz} \right) \chi^l_1+
\left( {z^2-l^2 \over z^2 } \right) \chi^l_1=0 . \eqno(16)$$
This is none other than Bessel's equation of order $l$. By regularity
at the origin, the solution is
$$
\chi^l_1=c_l J_l(\omega r) . \eqno(17)$$
This together with the second equation implies
$$
\chi^l_2=c_l J_{l+1} (\omega r). \eqno(18)$$
By a similar argument, $\psi^l_1$ and $\psi^l_2$ satisfy Bessel's equations
of order $l+1$ and $l+2$ respectively and the internal solution is
$$
\left( \matrix{\chi \cr \psi \cr} \right) = \sum^\infty_{l= -\infty}
 \left( \matrix{ c_l J_l(\omega r) \cr i c_l J_{l+1} (\omega r) e^{i \theta}
\cr d_l J_{l+1}(\omega r) e^{i \theta}\cr
 i d_l J_{l+2}(\omega r) e^{2i \theta}\cr} \right) e^{il \theta} . \eqno(19)$$

\noindent{\bf (b) External Solution ($r > R$)}

Outside the string core, we decompose our wave functions into eigenfunctions
of the helicity operator. i.e.,
$$
\matrix{
 (\sigma \cdot \pi_R) \chi &= -i \sigma^j D^R_j \chi &= \pm k \chi \cr
   (\sigma \cdot \pi_L) \psi &= -i \sigma^j D^L_j \psi &= \pm k \psi .\cr}
\eqno(20)
$$
{}From eqn.(14), this gives
$$
\eqalign{
 (\omega \mp k) \chi &= f_e f \psi = m \psi \cr
 (\omega \pm k) \psi &= f_e f \chi = m \chi .\cr} \eqno(21) $$
Defining
$$\nu = l - \alpha_R \eqno(22)$$
and $z'=kr$, eqn.(20) yields
$$
\chi^l_2= \mp \left( {d \over dz'} - {\nu \over z'} \right) \chi^l_1 ,\eqno(23)
$$ where $ -$ ($+$) is taken for a positive (negative) helicity state.
Thus, the external solution is
$$
\left( \matrix{\chi \cr \psi \cr} \right) = \sum^\infty_{l=-\infty}
\left( \matrix{ Z_\nu (kr) \cr \pm i Z_{\nu+1} (kr) e^{i \theta} \cr
  B^{\pm} Z_\nu (kr) e^{i \theta} \cr \pm i  B^{\pm} Z_{\nu+1}
 (kr) e^{2i \theta}\cr} \right) e^{il \theta} . \eqno(24) $$
In the above, $B^{\pm} = {m \over \omega \pm k} $, the superscript
$\pm$ in $B$ denotes
the helicity and $\pm$ in the front of the second and fourth components
take the same sign as the helicity for $Z_\nu = J_\nu$, $N_\nu$ and
$H_\nu$ and opposite sign for $Z_\nu = J_{-\nu}$, $N_{-\nu}$ and $H_{-\nu}$.
Here $N_\nu$ and $H_\nu$ are Neumann and
outgoing Hankel functions respectively.
Note that it is $kr $ rather than $\omega r$ which appears in the arguments
of our functions because electrons are massive outside the core. Another point
to note is that whereas the second and third components of the internal
solutions satisfy Bessel's equation of the same order, the corresponding
components of the external solutions satisfy Bessel's equations of orders
$\nu +1$ and $\nu$
respectively. This relative shift in the order is due to the asymmetry
between left and right, i.e., $\alpha_L= \alpha_R +1$.

\noindent{\bf (c) Asymptotic Solution}

Consider performing a scattering experiment with an incoming plane wave of
positive helicity electrons. Since helicity is violated in the core, the
scattered wave consists of both positive and negative helicity components.
We find that, as $r \to \infty$, the external solution takes the form
$$
\sum^{\infty}_{l=- \infty} e^{il \theta} \left[ \left( \matrix{ (-i)^l J_l \cr
      i (-i)^l J_{l+1} e^{i \theta} \cr B^+ (-i)^l J_l e^{i \theta} \cr
    i B^+ (-i)^l J_{l+1} e^{2i \theta}\cr} \right) + {f_l e^{ikr} \over
 \sqrt r} \left( \matrix{ 1 \cr e^{i\theta} \cr B^+ e^{i \theta} \cr
   B^+ e^{2i \theta} \cr} \right) + {g_l e^{ikr} \over \sqrt r} \left(
 \matrix{1 \cr - e^{i \theta} \cr B^- e^{i \theta} \cr -B^- e^{2i \theta} \cr}
\right) \right] . \eqno(25)$$
It is easy to check that the second and third terms are the positive and
negative helicity components of the scattered waves respectively.

We divide the problem of matching the asymptotic wave functions into two
cases.

(i) For $\nu \geq 0$ or $\nu \leq -1$, we take $Z_\nu^1=J_\nu $ and $Z_\nu^2
= N_\nu$.
The external wave function is therefore

$$
   \left(
\matrix{ (a_l J_\nu &  +& b_l   N_\nu &  +& A_l J_\nu & +& B_l  N_\nu)
e^{il \theta}  \cr
 i( a_l J_{\nu +1} &+& b_l N_{\nu +1}&-& A_l   J_{\nu +1}&-& B_l  N_{\nu +1})
e^{i(l+1) \theta}  \cr
(a_l B^+ J_\nu & +& b_l B^+ N_\nu & +& A_l B^- J_\nu &+& B_l B^- N_\nu )
e^{i(l+1) \theta}  \cr
 i (a_l B^+ J_{\nu+1} &+& b_l B^+ N_{\nu +1}& -& A_l B^- J_{\nu +1} &-&
B_l B^- N_{\nu +1} )e^{i(l+2) \theta}  \cr} \right) \eqno(26) $$

Making use of the asymptotic large $x$ forms
$$\eqalignno{J_{\mu} (x)
& \sim \sqrt {2 \over \pi x} \cos \left( x- {\mu \pi \over 2} - {\pi
\over 4} \right) \cr
N_{\mu} (x) & \sim \sqrt {2 \over \pi x} \sin \left( x- {\mu \pi \over 2}
 - {\pi
\over 4} \right), & (27)\cr}$$
we match coefficients of $e^{il \theta} {e^{\pm ikr} \over \sqrt r} $ in
eqns.(25) and (26) to find
$$
\matrix{ \hfill e^{i \nu \pi /2} & (a_l &+ i b_l &+ A_l &+ i B_l )&=
   1 \hfill \cr
   \hfill e^{-i \nu \pi /2} & ( a_l &- ib_l &+ A_l &-i B_l)&=
  (-1)^l
 + (f_l + g_l) e^{i \pi /4} \sqrt {  2 \pi k} \hfill \cr
  \hfill e^{i (\nu+1) \pi /2} &(ia_l &- b_l &-i A_l &+  B_l )&=
  -1 \hfill \cr
 e^{-i (\nu+1) \pi /2} &
( ia_l &+b_l &-i A_l &- B_l)&=
  (-1)^l +
 (f_l - g_l) e^{i \pi /4} \sqrt {  2 \pi k}, \cr} \eqno(28) $$
from which we deduce
$$\eqalignno{ A_l &= -i B_l \cr
              a_l &= -i b_l + e^{-i \nu \pi /2} \cr
              g_l &= e^{-i( \pi /4+ \nu \pi /2) }  \sqrt {1 \over 2 \pi k}
                     (-2i B_l) .&(29) \cr} $$

(ii) For $-1 < \nu < 0$, taking $Z_\nu^1= J_\nu$ and $Z_\nu^2= J_{-\nu}$, the
external wave function is

$$
   \left(
\matrix{ (a_l J_\nu &  +& b_l   J_{-\nu} &  +& A_l J_\nu & +& B_l  J_{-\nu})
e^{il \theta}  \cr
 i( a_l J_{\nu +1} &-& b_l J_{-\nu -1}&-& A_l   J_{\nu +1}&+& B_l  J_{-\nu -1})
e^{i(l+1) \theta}  \cr
(a_l B^+ J_\nu & +& b_l B^+ J_{-\nu} & +& A_l B^- J_\nu &+& B_l B^- J_{-\nu} )
e^{i(l+1) \theta}  \cr
i (a_l B^+ J_{\nu+1} &-& b_l B^+ J_{-\nu -1}& -& A_l B^- J_{\nu +1} &+&
B_l B^- J_{-\nu -1} )e^{i(l+2) \theta}  \cr} \right) \eqno(30) $$

We proceed as before and find
$$
\eqalignno{A_l &= - e^{-i \nu \pi} B_l \cr
           a_l &= - e^{-i \nu \pi} b_l + e^{-i \nu \pi /2} \cr
           g_l &= \sqrt {1 \over 2 \pi k} e^{-i (\nu \pi /2 + \pi /4)} 2i
    \sin(\nu \pi) B_l. &(31)}$$

Note that
$${d \sigma \over d \theta }\biggr|_{+ \to -} = \sum_l |g_l|^2
. \eqno(32)$$

\noindent{\bf (d) Matching at $r=R$}

We have obtained the solutions inside and outside the core in (a) and (b).
Now we match them at $r=R$. Because of the difference in the masses in the
two regions and the discontinuous distribution of the string flux, there is
a discontinuity in the first derivatives of the wave functions. Nevertheless,
the wave functions themselves are continuous at $r= R$. This is the matching
condition that we will use.\refmark{\Alf,\Per} Once again there are two cases.

(i) $ \nu \geq 0 $ or $ \nu \le -1$:
Substituting (29) into (26) and matching it with the internal solution
in eqn.(19), we obtain
$$
\matrix{
-i H_\nu b_l& -&  i H_\nu B_l& =&  J_l c_l &-&
 e^{-i\nu \pi /2} J_\nu \cr
 -i H_{\nu+1} b_l &+&i H_{\nu+1} B_l& =& J_{l+1} c_l &-&
e^{-i\nu \pi /2} J_{\nu+1} \cr
 -iB^+ H_\nu b_l &-&iB^- H_\nu B_l &=& J_{l+1} d_l &-&
e^{-i\nu \pi /2}B^+ J_\nu \cr
-i B^+H_{\nu+1} b_l &+&i B^- H_{\nu+1} B_l &=& J_{l+2} d_l
 &-&e^{-i\nu \pi /2} B^+ J_{\nu+1}. \cr}\eqno(33) $$
In deriving the above equations, we have used the definition of outgoing
Hankel function:
$H_\mu= J_\mu+i N_\mu .$
Solving eqn.(33), we find
$$
B_l= { \Delta_B \over \Delta} \eqno(34)$$
where
$$ \Delta_B=B^+ e^{-i \nu \pi /2} ({ 2 \over \pi k R})
 ( J_{l+1}^2 -J_l J_{l+2})
\eqno(35)$$ and
$$
\Delta= (B^- - B^+) J_{l+1} (J_l H^2_{\nu +1} - J_{l+2} H^2_{\nu}) -
        (B^- + B^+) (J_{l+1}^2 - J_l J_{l+2}) H_\nu H_{\nu +1} \eqno(36)$$
where $H_\nu$ is outgoing Hankel functions. Use has been made
of the Wronskian formula
$J_{\nu+1}(x) N_\nu(x) - J_\nu(x) N_{\nu+1}(x) = {2 \over \pi x}$ in the
derivation of eqn.(35).

Now we consider the regim\'e $\omega \sim k \sim m$ and $k R \ll 1$ and perform
small $kR$ approximation:
$$ J_\mu  \sim O([kR]^\mu),\qquad H_\mu, N_\mu \sim O([kR]^{-|\mu|}).
\eqno(37)$$
It is straightforward, but tedious to show that
$$
B_l \propto \cases{ (kR)^{2 \nu+ 2} & $ \nu \geq 0, l \geq 0$ \cr
                    (kR)^{2 \nu} & $\nu \geq 0, l <0$ \cr
                    (kR)^{-2 \nu -2} & $\nu \le -1, l \geq -1$ \cr
                    (kR)^{-2 \nu} & $\nu \le -1, l < -1.$  \cr} \eqno(38)$$

Note that the cross section may still be logarithmically suppressed when the
exponent in the suppression factor appears to be zero.

(ii) $-1 < \nu <0$

Substituting eqn.(31) into (30) and matching it with the internal solution
in eqn.(19), we obtain the following equations.
$$
\matrix{P(& b_l& +& B_l&)= & J_l c_l&- &
  e^{-i \nu \pi /2}J_\nu \cr
Q(& b_l& - & B_l&) = &  J_{l+1} c_l& -&
 e^{-i \nu \pi /2}J_{\nu+1} \cr
P(&  B^+ b_l& + & B^- B_l&)= & J_{l+1} d_l &- &
   e^{-i \nu \pi /2} B^+ J_\nu \cr
Q(&  B^+ b_l &- & B^- B_l&)= & J_{l+2} d_l &- &
   e^{-i \nu \pi /2} B^+ J_{\nu+1} \cr}, \eqno(39) $$
where $P$ denotes $ -e^{-i \nu \pi } J_\nu+ J_{-\nu} $
and $Q$ denotes $ -e^{-i \nu \pi } J_{\nu+1}- J_{-\nu-1} $.
Solving (39), we find
$$
B_l= {\Delta'_B \over \Delta'} \eqno(40)$$
where
$$
\Delta'_B={2  B^+ \sin(\nu \pi) e^{- i\nu \pi /2} \over \pi k R}
 (J_l J_{l+2} -J^2_{l+1} ) \eqno(41)$$
and
$$
\Delta'=(B^- -B^+) J_{l+1} (J_{l+2} P^2 - J_l Q^2) + (B^- + B^+)
( J^2_{l+1} -J_l J_{l+2}) PQ . \eqno(42)$$
Use has also been made
of the Wronskian formula $ J_\nu J_{-\nu -1} + J_{- \nu} J_{\nu+1}=
- {2 \sin(\nu \pi) \over \pi x} $ in deriving eqn.(41).

We consider the regim\'e $\omega \sim k \sim m$ and $kR \ll 1.$
A straightforward calculation shows that
$$
B_l \propto \cases{  (kR)^{2 \nu +2} &$ l \geq 0$ \cr
                     (kR)^{-2 \nu} & $l \leq -2$ \cr
                      1             &$ l= -1$ .\cr} \eqno(43)$$

We see immediately that when the last case occurs, the mode $l=-1$
($-1 < \nu <0$) dominates the contribution from all other modes, and is
of order 1. In that case, the helicity flip process is maximally enhanced with
a cross section of order $1/ m$, where
$m$ is the mass of the incoming particle.
Recalling that $\nu= l- \alpha_R, $ we see that this occurs precisely
when $-1< \alpha_R <0$ (and thus $0 < \alpha_L <1. $) For electrons
$\alpha_R= -2 \sin^2\theta_w $ and the condition reduces to $ 0 < \sin^2
\theta_w < 1/2 .$

Let us consider the changes in the helicity flip cross section as $\sin^2
\theta_w$ increases from $0$ to $1$ in the regim\'e $\omega \sim k \sim m$
and $kR \ll 1$.

(1) $\sin^2 \theta_w =0$ ($ \alpha_R=0$)

(2) $ 0 < \sin^2 \theta_w <{1 \over 2}$ ($-1 < \alpha_R <0$)

(3) $  \sin^2 \theta_w ={1 \over 2}$ ($ \alpha_R= -1$)

(4) $ {1 \over 2} < \sin^2 \theta_w < {3 \over 4}$ ($-1.5 < \alpha_R <-1$)

(5) $ \sin^2 \theta_w = {3 \over 4}$ ($\alpha_R=-1.5$)

(6) $ {3 \over 4} < \sin^2 \theta_w < 1 $ ($-2 < \alpha_R <-1.5$)

(7) $ \sin^2 \theta_w =1 $ ($\alpha_R= -2$)

Before embarking on a discussion about the various cases for electrons, we
would like to remark that the results for down quarks are similar.
It is still true
that $\alpha_L= \alpha_R+1$. The only difference is that
$\alpha_R= -{2
\over 3} \sin^2 \theta_w $ for d quarks. Therefore, there are just two cases.
If $\sin^2 \theta_w =0,$ the result is the same as in case (1) for electrons
and the helicity flip scattering has, up to normalization, an Everett's
cross section. (Cf. case (1) below.)
If $0<\sin^2 \theta_w \leq 1,$ there is a maximal enhancement
and the cross section per unit length $ \sim 1 /m_d$. (Cf. case (2) below.)
Now we turn to electrons.

(1) For $\sin^2 \theta_w =0$ ($ \alpha_R=0$), the $l=\nu=-1$ mode dominates
and from eqns. (29), (32) and (34)-(36), the differential
cross section per unit length
$${d \sigma \over d \theta} \sim {1 \over k {\ln}^2 (kR)}, \eqno(44)$$
which is, up to normalization, the cross section obtained by
Everett\Ref\Eve{A.
E. Everett, Phys. Rev. {\bf D 24}, 858 (1981).}
for the scattering of scalar particles off cosmic strings with integral
magnetic fluxes.

(2) For $ 0 < \sin^2 \theta_w <{1 \over 2}$ ($-1 < \alpha_R <0$),
from eqns. (31), (32) and (43), the electron
helicity flip process has a differential cross section
$$ {d \sigma \over d \theta} =O( m_e^{-1})\eqno(45)$$ which is dominated by
the mode $-1<\nu<0$ ($l=-1$) and is thus independent of angle. In
this case, the helicity flip process remains unsuppressed as
$R \to \infty$ with $k$ held fixed. Note that
this maximal amplification occurs for
a continuous range of values of the parameter $\sin^2 \theta_w$. This is
in contrast with an analogous calculation on baryon number violating
processes due to cosmic strings which exhibit unsuppressed cross section
for only discrete values of fluxes.\refmark{\Alf,\Per}

(3) For $  \sin^2 \theta_w ={1 \over 2}$ ($ \alpha_R= -1$), the $l=-1,
\nu=0$ mode swamps contributions from all other modes. Eqns. (29),
(32) and (34)-(36) together implies that the cross section is
of the same order as in case (1).

(4) For ${1 \over 2}< \sin^2 \theta_w < {3 \over 4}$ ($-1.5 < \alpha_R <-1$),
the dominant mode is $ 0< \nu <0.5$ ($l=-1$). From eqns. (29), (32)
and (38), the differential cross section
is given by
$$ {d \sigma \over d \theta} \sim k^{-1} (kR)^{4 \nu} =
 k^{-1} (kR)^{4(2 \sin^2 \theta_w -1)}, \eqno(46)$$

(5) For $ \sin^2 \theta_w = {3 \over 4}$ ($ \alpha_R = -1.5$), the two modes
$l=-1$ and $-2$ give comparable contributions and we obtain from eqns.
(29), (31), (32), (38) and (43) that
$$ {d \sigma \over d \theta} \sim {1 \over k (kR)^2} |1+ Ce^{i \theta}|^2 .
\eqno(47)$$

(6) For ${3 \over 4} <\sin^2 \theta_w <1$ ($ -2 <\alpha_R < -1.5$), we
need to consider the contribution from the $l=-2$ mode only and obtain
from eqns. (31), (32) and (43) that
$$ {d \sigma \over d \theta} \sim
 k^{-1} (kR)^{8(1- \sin^2 \theta_w )}, \eqno(48)$$

(7) For $\sin^2 \theta_w =1$ ($\alpha_R=-2$), the $l=-2, \nu=0$ mode
will dominate and the differential cross section can be deduced from
eqns. (29), (32) and (34)-(36):
$${d \sigma \over d \theta} \sim {1 \over k \ln^4 (kR)}. \eqno(49)$$
Note that the exponent of the logarithmic term is four,
whereas in cases (1) and (3) it is two.
We note on passing that case (7) corresponds to semi-local strings,
where the $SU(2)$ gauge charge is set to zero.

The most prominent feature of our result is the presence of a plateau: For
$\sin^2 \theta_w$ between $0$ and $1/2$, we have maximal enhancement. Is there
any heuristic way of understanding its origin? In Ref. 6, the cross section
for a baryon number violating process was derived using first order
perturbation theory in quantum field theories. The transition matrix element
between an initial state $\ket{\psi}$ and a final state $\ket{\psi'}$ is given
by
${\cal A} = \langle \psi'| \int d^4 x {\cal L}_I (x) \ket{\psi}.$
The computation is divided into two steps. Firstly, we evaluate ${\cal A}$
using free fermion spinors, resulting in the ``geometric'' cross section.
In the second step, we solve the Dirac equation with the appropriate boundary
conditions to determine the amplitude of the spinor at the core radius R and
define the amplification factor $A$ as the ratio of the amplitude of the
actual spinor to that of a free spinor. Since the cross section is
proportional to ${\cal A}^2$ and since ${\cal A}$ involves two spinors, the
catalysis cross section is enhanced by a factor $A^4$ over the geometric
cross section. Now we attempt a similar discussion for helicity flip
due to electroweak strings. The difficulty of such an approach lies in
the decomposition of the Hamiltonian into helicity conserving and
helicity violating parts. The point is that the Yukawa coupling between
the Higgs field and electrons, apart from giving rise to helicity violation,
also makes electrons massive and it seems difficult to separate these
two effects.
The simplest way out is to consider another object instead, namely
the commutator of the helicity operator with the Hamiltonian. This object
is clearly proportional to the transition matrix element that we are interested
in.\foot{We thank Ming Lu and Piljin Yi for helpful discussions about this.}
{}From eqn.(11) we see that this commutator couples the first component of
a spinor with the fourth and the second with the third, etc. In the same
spirit as in Ref.6, we compute the coefficients $a_l$, $b_l$, $A_l$ and $B_l$
of our wave function. For the region $0<\sin^2 \theta_w <1/2$, we find that
the mode $ -1 <\nu <0$ have all coefficients of order unity. It is a simple
matter to check that, at the core radius, the first and third components
of the initial ($+$ helicity) wave function are enhanced by a factor
$(kR)^{\nu}$ and the second and fourth by $(kR)^{-\nu-1}$. A similar analysis
holds for the final ($-$ helicity) state. Now we have the interesting result
that all components are amplified by factors very sensitive to the fractional
flux of the string, but the first and third components have a
different amplification factor from that of the second and fourth
such that when we take the product of the amplification factors, we get
an enhancement factor of $(kR)^{-1}$, which is independent of $\alpha_R$.
This is the origin of the plateau.

\chapter{Concluding Remarks}

(1) We work in the regim\'e $\omega \sim k \sim m_e$
and $kR \ll 1$.
Using a partial wave decomposition, we show that for $0 <\sin^2
\theta_w < {1/2},$ electrons scattering off an electroweak string
have a helicity flip cross section
(per unit length) of order $m_e^{-1}$. This huge
cross section is due to an amplification of the fermionic wave function
at the core.  Within this region of the parameter space,
it is found that one partial wave (the
mode $-1 <\nu <0$) dominates the contributions from all other modes,
giving an angle independent differential cross section (per unit length)
$\propto m^{-1}_e.$

(2) Whereas baryon number violating processes are maximally enhanced only
for discrete values of the fractional flux, our results show that electroweak
strings have maximal amplified helicity flip scattering amplitude for a
continuous region of the parameter space $0<\sin^2\theta_w< 1/2$. This is
due to the asymmetry between left and right and a subsequent
delicate cancellation of the dependence of the overall amplification factor
on the Weinberg angle:
We consider the commutator between the helicity
operator and the Hamiltonian. This commutator gives a coupling between
the first and fourth components as well as between the second and third
components
of the spinor. By computing the coefficients of the wave function,
one observes that the first and third components are amplified at a factor
which is very sensitive to the string flux and is different from the
amplification factor for the second and fourth components. However, when
we take the product of these two amplification factors to obtain the total
amplification factor, we find it to be independent of the string flux, thus
resulting in maximal enhancement for a continuous region of the parameter
space.

(3) The case $0 \leq \sin^2\theta_w \leq 1/2$
has been discussed in a revised version of Ref. 14.
Can one rederive their results from our discussion? The answer is affirmative.
For $0<\sin^2\theta_w< 1/2$ and $\omega R$, $kR \ll 1$, one can
deduce from
eqns. (31), (32) and (40)-(42) that
$${d \sigma \over d \theta} \approx {2 \over \pi k } \left(
{m^2 \over \omega (\omega +k) }\right)^2 \sin^2 \pi \alpha_R \eqno(50).$$
(Here we have reinstated the mild $\sin^2 \pi \alpha_R$ dependence
that we have ignored in section 3.) In the limit $k \ll m$,
this gives
$$ {d \sigma \over d \theta} \approx {2 \over \pi k }
 \sin^2 \pi \alpha_R . \eqno(51)$$
In the opposite limit $k \gg m$,
$$ {d \sigma \over d \theta} \approx {1 \over 2 \pi k} \left(
{m \over k} \right)^4  \sin^2 \pi \alpha_R . \eqno(52)$$
We note that the vanishing of the differential cross section
in the massless limit can be deduced directly from eqn.(11).
Similarly, for $\sin^2 \theta_w =0 $ or $1/2$, one deduces from eqns.(29),
(32) and (34)-(36) that
$${d \sigma \over d \theta} \approx {\pi \over 8 k } \left(
{m^2 \over \omega (\omega +k) }\right)^2 {1 \over \ln^2 kR} \eqno(53).$$
For  $k \ll m$, this gives
$${d \sigma \over d \theta} \approx {\pi \over 8 k }
 {1 \over \ln^2 kR} \eqno(54).$$ For $k \gg m$, this gives
$${d \sigma \over d \theta} \approx {\pi \over 32 k } \left(
{m \over k }\right)^4 {1 \over \ln^2 kR} \eqno(55).$$
These results are in good agreement with those of Ref. 14.
(The authors
gave zero as the final answer to the case $k \gg m$, but it is
clear from their arguments that they had neglected $\left( {k /m} \right)^4$
terms.)

(4) The cross section for other values of $\theta_w$ are also computed. In
particular, a semi-local string is none other than an
electroweak string with $\sin^2 \theta_w=1$.

(5) For axion strings, the covariant derivatives in eqn.(11) should be
replaced by partial derivatives. Therefore, helicity is violated outside
the core and it makes no sense to talk about helicity conserving and
helicity flip cross sections in this context.

(6) It would be interesting to investigate the relevance of this work
to electroweak baryogenesis. Such scenarios are highly testible.

\bigskip
We gratefully acknowledge useful discussions with Ming Lu,
Piljin Yi, and particularly John Preskill. We also thank A. P. Martin for
sending us a revised version of Ref. 14 and pointing out some errors
in an earlier version of this manuscript.

\refout

\end